\documentclass[twocolumn,prl,showpacs]{revtex4}
\usepackage{graphics}

\begin{document}
                                                         
\title{Comment on ``Loss of Superconducting Phase Coherence in
 ${\rm YBa}_2{\rm Cu}_3{\rm O}_7$ Films:
 Vortex-Loop Unbinding and Kosterlitz-Thouless Phenomena''}
\author{Kateryna \surname{Medvedyeva}}
\author{Beom Jun \surname{Kim}}
\author{Petter \surname{Minnhagen}}
\affiliation{Department of Theoretical Physics, Ume{\aa} University,
901 87 Ume{\aa}, Sweden}


\maketitle

Recently, K\"{o}tzler {\em et al.}~\cite{kotzler} measured 
the frequency-dependent conductance for ${\rm YBa}_2{\rm Cu}_3{\rm O}_7$ 
and interpreted their results as evidences that the decay of the 
superfluid density is caused by a 3D vortex loop  proliferation mechanism 
and a dimensional crossover when the correlation length
$\xi_c$ along the $c$ axis becomes comparable to the sample 
thickness $d$.  In this Comment, we show that  the complex conductance
data presented in Ref.~\onlinecite{kotzler} have characteristic key features
not compatible with their
analysis, which are instead described by the existing 
phenomenology of 2D vortex fluctuation
in Ref.~\onlinecite{minnhagen:rev} associated with a partial
decoupling of ${\rm Cu}{\rm O}_{\rm 2}$-planes. It is also argued that the
suggested dimensional crossover makes the fluctuations stronger,
and accordingly shifts the transition towards lower temperature. 
This appears to be opposite to the conclusion in Ref.~\onlinecite{kotzler}.

\begin{figure}
\centering{\resizebox*{!}{5.5cm}{\includegraphics{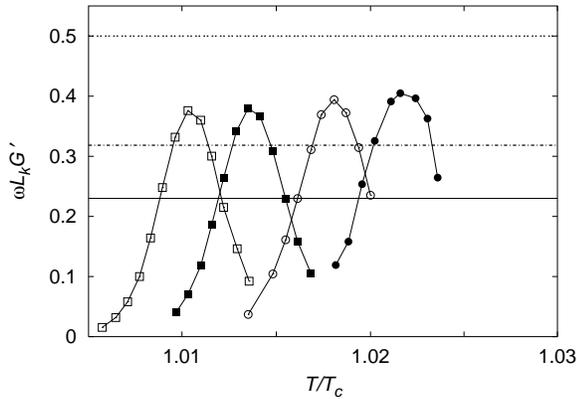}}}
\caption{Dissipative part
  $\omega L_k(T) G'$ 
  vs $T/T_c$ from Ref.~\onlinecite{kotzler}. The curves correspond to
   $\omega = 30$mHz, $3$Hz, $1$kHz, and $100$kHz (from left to right).}
\label{fig:1}
\end{figure}

In Fig.~1, $G'$ in the complex conductance $G=G'+iG''$ from Fig.~2 in
Ref.~\onlinecite{kotzler}
is replotted as $\omega G'L_{k}$, with the kinetic inductance
$L_{k}(T)$
extracted in Ref.~\onlinecite{kotzler}. In 
Ref.~\onlinecite{kotzler}, it is assumed that 
$\omega L_k(T)G(\omega,T)= S(\omega/\omega_0)$ with a scaling function
$S = S' + iS''$ satisfying
\begin{equation} \label{S}
  S''= 1/[1+(\omega/\omega_0)^{-\alpha}]
\end{equation}
and $S'$ is obtained from the Kramer-Kronig relation~\cite{minnhagen:rev}.
K\"{o}tzler {\em et al.} find that their
data can be characterized by $\alpha\approx 0.7$, which corresponds to 
the heights of the dissipation peaks equal to 0.23 (solid 
line in Fig.~1). However the data exceed this
value by more than $40\%$, and the peak height shows a systematic
increase with increasing frequency, whereas Eq.(\ref{S}) predicts
constant peak heights: These observations imply that the
scaling function~(\ref{S}) cannot properly describe the experimental data.
The 2D vortex explanation, on the other hand, predicts that 
the peak heights should systematically increase with $\omega$ 
and should always be between $1/\pi\approx 0.32$ and 0.5 
(dashed and dotted lines in the Fig.~1)~\cite{minnhagen:rev,festin}. 
The lower value $1/\pi$ corresponds to the Minnhagen response form 
[given by $\alpha=1$ in Eq.~(\ref{S})] associated with bound
vortex-antivortex pairs dominating close to the Kosterlitz-Thouless (KT)
transition. The higher value $1/2$ corresponds to the Drude 
response [given by $\alpha=2$ in Eq.~(\ref{S})], dominated by
abundant free vortices well above the KT transition. 
Both free vortices and bound pairs are present above the KT
transition with the proportion of free vortices increasing with $T$. This
explains why the peak heights increase with
$T$~\cite{minnhagen:rev,festin}.

K\"{o}tzler {\em et al.} link their proposed mechanism to an apparent 
sample independent onset of
deviation at about 35(nH)$^{-1}$ for the inductive data taken at 1kHz from the
inferred 3D $L^{-1}_k$. The relevant scale where vortex type fluctuations
become important may be estimated from the KT transition criteria $L^{-1}_k=8.2
\times 10^{-2} T$ (KnH)$^{-1}$ giving $L^{-1}_k\approx 7$(nH)$^{-1}$ for the
samples in Ref.~\onlinecite{kotzler}. This places the KT-transition
(which in principle always can be associated with a large enough sample
of finite thickness) very close to the inferred 3D transition. 
Although 35(nH)$^{-1}$ is larger, it is
still of the same order of magnitude. This suggests that the deviation at
1kHz occurring below the inferred
critical $T_c$ (as well as  below the inferred KT-transition) may
indeed be caused by vortex loops. However, in such a
scenario the deviation from the inferred zero frequency $L^{-1}_k$ towards
higher values is due to the {\em finite frequency}~\cite{minnhagen:rev}. This
deviation would then vanish in the limit of small frequency (small in the sense
that the dissipation peaks in Fig.~1 cease to move to the left) implying a {\em
frequency dependent} onset of deviations. Any change of the 3D $L^{-1}_k$
caused by a dimensional crossover ($\xi_c \gtrsim d$) will always increase
fluctuations and cause deviation in the limit of zero frequency to the {\em
left} instead of to the {\em right}  of the inferred
3D $L^{-1}_k$.

In our scenario the deviation occurs when $\omega/\omega_0$ becomes
large which always happens for a $T$ below $T_c$ because
$\omega_0\rightarrow 0$ at $T_c$. The reason is that the larger
$\omega/\omega_0$ the more of the larger vortex loops become
ineffective in renormalizing $L_k^{-1}$, causing an increase from
the $\omega=0$-value \cite{minnhagen:rev}. The universality found in
Ref.\onlinecite{kotzler} implies that $\omega_0\propto L_k^{-1}$ which is
similar to what happens in the 2D XY model \cite{jonsson}. The point to
note is that in order for the data to deviate to the right of the
$\omega=0$-data the fluctuations involved in the $\omega=0$-data
should be the ones that are made ineffective by the finite
frequency. Since the $\omega=0$-data obeys the 3D critical scaling
the relevant fluctuations would have to be the usual vortex
loops. Consequently the data is well explained without invoking
a vortex blow out involving an additional $\xi_c/d\approx 1$ condition.


\end{document}